\newcommand{\ourwork}{TransFetch}
\documentclass[sigconf]{acmart}
\AtBeginDocument{%
  \providecommand\BibTeX{{%
    \normalfont B\kern-0.5em{\scshape i\kern-0.25em b}\kern-0.8em\TeX}}}

\setcopyright{acmcopyright}
\copyrightyear{2022}
\acmYear{2022}
\setcopyright{rightsretained}
\acmConference[CF'22]{19th ACM International Conference on Computing Frontiers}{May 17--19, 2022}{Torino, Italy}
\acmBooktitle{19th ACM International Conference on Computing Frontiers (CF'22), May 17--19, 2022, Torino, Italy}\acmDOI{10.1145/3528416.3530236}
\acmISBN{978-1-4503-9338-6/22/05}

\usepackage{subfig}
\usepackage{microtype}
\usepackage{amsmath}

\usepackage{amssymb}

\usepackage{newtxmath}
\usepackage{amsfonts}
\usepackage{graphicx}

\usepackage{url}

\begin{document}

\title[Fine-Grained Address Segmentation for Attention-Based Variable-Degree Prefetching]{Fine-Grained Address Segmentation for\\Attention-Based Variable-Degree Prefetching}


\author{Pengmiao Zhang}
\orcid{0000-0002-5411-3305}
\affiliation{%
  \institution{University of Southern California}
  \city{Los Angeles}
    \state{California}
  \country{USA}
  \postcode{90089}
}
\email{pengmiao@usc.edu}

\author{Ajitesh Srivastava}
\affiliation{%
  \institution{University of Southern California}
  \city{Los Angeles}
  \state{California}
  \country{USA}
  \postcode{90089}}
\email{ajiteshs@usc.edu}

\author{Anant V. Nori}
\affiliation{%
  \institution{Processor Architecture Research Lab, Intel Labs}
\city{Bangalore}
  \country{India}
  }
\email{anant.v.nori@intel.com}

\author{Rajgopal Kannan}
\affiliation{%
  \institution{US Army Research Lab-West}
  \city{Los Angeles}
  \state{California}
  \country{USA}
 }
\email{rajgopal.kannan.civ@army.mil}
 
\author{Viktor K. Prasanna}
\affiliation{%
  \institution{University of Southern California}
  \city{Los Angeles}
  \state{California}
  \country{USA}
}
\email{prasanna@usc.edu}

\renewcommand{\shortauthors}{Pengmiao, et al.}

\begin{abstract}

Machine learning algorithms have shown potential to improve prefetching performance by accurately predicting future memory accesses. Existing approaches are based on the modeling of text prediction, considering prefetching as a classification problem for sequence prediction. However, the vast and sparse memory address space leads to large vocabulary, which makes this modeling impractical. The number and order of outputs for multiple cache line prefetching are also fundamentally different from text prediction.

We propose TransFetch, a novel way to model prefetching. To reduce vocabulary size, we use fine-grained address segmentation as input. To predict unordered sets of future addresses, we use delta bitmaps for multiple outputs. We apply an attention-based network to learn the mapping between input and output. Prediction experiments demonstrate that address segmentation achieves 26\% - 36\% higher F1-score than delta inputs and 15\% - 24\% higher F1-score than page \& offset inputs for SPEC 2006, SPEC 2017, and GAP benchmarks. Simulation results show that TransFetch achieves 38.75\% IPC improvement compared with no prefetching, outperforming the best-performing rule-based prefetcher BOP by 10.44\% and ML-based prefetcher Voyager by 6.64\%.

\end{abstract}

\ccsdesc[500]{Computer systems organization~Processors and memory architectures}

\ccsdesc[500]{Computer systems organization~Neural networks} 

\ccsdesc[100]{Information systems~Data mining}

\keywords{prefetching, machine learning, attention, address segmentation}


\maketitle




\section{Introduction}
Memory latency is becoming an overwhelming bottleneck in computer performance due to the "memory wall"~\cite{wulf1995hitting,carvalho2002gap} problem, especially with the advent of GPUs~\cite{nvidia2017v100}, TPUs~\cite{jouppi2017datacenter}, and heterogeneous architectures~\cite{hazarika2020survey, reuther2019survey} that accelerate computation. Prefetching is critical in reducing program execution time and improving instructions per cycle (IPC) by hiding the latency. It looks at patterns of memory accesses sequences and uses the past information to forecast the near future accesses so as to start fetching the data before the miss occurs~\cite{dubois2012parallel, vander1997caches}. Existing prefetchers are mainly heuristic, predicting via pre-defined rules, based on the observation from the locality of references~\cite{smith1978sequential,jouppi1990improving,hur2006memory,palacharla1994evaluating,chen1995effective,farkas1997memory,somogyi2006spatial,ishii2011access,johnson1997run,kumar1998exploiting,lin2001filtering,chen2004accurate,pugsley2014sandbox,michaud2016best,shevgoor2015efficiently,wenisch2008temporal,chilimbi2001efficient,joseph1997prefetching,nesbit2004ac,solihin2002using,hu2003tcp,chou2007low,jain2013linearizing,bakhshalipour2018domino,wu2019efficient,wu2019temporal}. 
With the rise of new workloads, such as graph analytics~\cite{murphy2010introducing, beamer2015gap,samsi2018graphchallenge}, data mining~\cite{tsai2015big,kambatla2014trends}, and AI applications~\cite{ignatov2018ai,wahl2018artificial,vaishya2020artificial}, rule-based prefetchers are not powerful enough to adapt to the increasingly irregular, indirect, and complex memory access patterns.

Machine learning-based data prefetchers are gaining increasing attention to pursue higher performance for memory access prediction~\cite{hashemi2018learning, srivastava2019predicting,mempatterns} and prefetching~\cite{peled2015semantic,peled2018neural, zeng2017long, braun2019understanding,srivastava2019predicting,shi2021hierarchical}. Prefetching is commonly modeled as classification for sequence prediction~\cite{hashemi2018learning, srivastava2019predicting,mempatterns,peled2018neural,braun2019understanding,zhang2021c,zhang2020raop,shi2021hierarchical}, which is analogous to the problem setting of text prediction~\cite{hashemi2018learning} in natural language processing (NLP)~\cite{nadkarni2011natural}. However, this analogy is not perfect. 
First, the unique memory addresses for an application can be tens of millions, which is orders of magnitude larger than natural language vocabulary and exceeds the capability of machine learning models. This problem is known as \textit{class explosion}~\cite{shi2021hierarchical}. Existing approaches partly solve this problem by working on memory access address deltas~\cite{hashemi2018learning, srivastava2019predicting,mempatterns} or splitting an address by page and offset~\cite{shi2021hierarchical}. Second, \textit{tokenization}~\cite{webster1992tokenization}, as a commonly used technique in NLP that maps a meaningful word into nonsensitive numerical data for model processing, is also borrowed by existing ML-based prefetching models for preprocessing. 
Tokenization results in extra storage to save the mapping tables (token dictionaries) in hardware implementation, but the cost has been neglected by previous works~\cite{hashemi2018learning, srivastava2019predicting,mempatterns,peled2018neural,braun2019understanding,zhang2021c,zhang2020raop,shi2021hierarchical}. Third, unlike text prediction with a ground truth of future words in a fixed order, in prefetching there is no ground truth of a certain memory address that should be prefetched. This is know as the \textit{labeling} problem~\cite{shi2021hierarchical}. Any access address following the current access could be a potential label. For a prefetcher that can prefetch multiple blocks for each trigger (prefetch degree $>1$), the order of predicted block addresses for one prefetch is also insignificant. Lastly, the \textit{latency} overhead of ML-based prefetcher is also ignored under this setting. LSTM (Long short-term memory)~\cite{hochreiter1997long} is a commonly used prediction model~\cite{srivastava2019predicting,srivastava2020memmap,zhang2020raop,zhang2021c,shi2021hierarchical,mempatterns} due to its advantage in sequence modeling. However, the recurrent structure of LSTM is hard to be parallelized and the inference latency increases linearly with the input time step length. Recent success of attention-based models, e.g. the Transformer~\cite{vaswani2017attention}, provides insight into solving this problem by the virtue of high parallelizability.

To solve the problem of \textit{class explosion, tokenization, labeling}, and \textit{latency}, we propose~\ourwork~(\underline{Trans}former for pre\underline{fetch}ing), an attention-based prefetcher that supports variable-degree prefetching. We model prefetching as a multi-label classification problem. To overcome class explosion for memory address input, we propose an address segmentation method to reduce the vocabulary without information loss. It avoids tokenization as the processed value can be directly fed into a neural network. For labeling, we use delta bitmaps that collects unordered sets of future deltas to the current address, which paves the way for multiple block (cache line) prefetching. In inference, an optimal threshold that maximizes the F1-score is adapted to adjust the prefetch degrees (the number of blocks to be prefetched) and balance prefetch aggressiveness. We apply a powerful and embarrassingly parallelizable attention-based network to learn the mapping between the input segmented addresses and the delta bitmap labels. The model also supports incorporation of more context features (program counters, page distances) to enhance the prediction performance. Besides, we further offset the model inference latency by artificially introducing estimated latency in training and then performing distance prefetching.

Overall, our main contributions are:
\begin{itemize}
    \item We propose~\ourwork, an ML-based prefetcher that models prefetching as multi-label classification. Our model uses address segmentation for input, delta bitmap for labeling, attention-based network with context enhancement for prediction, optimal-threshold confidence throttling mechanism for variable-degree prefetching, and distance prefetching for hiding inference latency.
    
    \item We demonstrate the effectiveness of address segmentation, attention-based network, and context enhancement in prediction experiments. Results show that address segmentation achieves 26\% - 36\% higher F1-score than delta inputs and 15\%- 24\% higher F1-score than page \& offset inputs. Attention-based model achieves 10\% - 13\% higher F1-score than LSTM and Temporal Convolutional Networks (TCN)~\cite{lea2017temporal}. Context enhancement raises the F1-score by 3.1\% - 9.1\%.
    
    \item We evaluate the performance of~\ourwork~using accuracy, coverage, and IPC improvement. Results show that our method achieves 88.56\% prefetch accuracy and 60.54\% prefetch coverage. It improves IPC by 38.75\% compared with no prefetching, outperforming the best-performing rule-based prefetcher BOP by 10.44\%, and ML-based prefetcher Voyager by 6.64\%. 
\end{itemize}

\section{Background and Related Work}
In this section, we provide background for data prefetching, 
attention mechanism,
along with the related prior works.

\subsection{Data Prefetching}
A prefetching process is a form of speculation that aims to predict the future data addresses and fetch the data before it is needed. 
Prefetch degree is the number of fetching blocks for each prefetching operation, which indicates the aggressiveness of a prefetcher. While a higher degree is likely to bring more useful data into cache, it may introduce cache pollution due to wrong predictions.

\noindent\textbf{Rule-based prefetching.} Traditional prefetchers learn from pre-defined rules, usually exploiting spatial or temporal localities. For example, Spatial Memory Streaming (SMS)~\cite{somogyi2006spatial} prefetcher identifies code-correlated spatial patterns to predict future accesses. Spatial prefetcher BOP~\cite{michaud2016best} and VLDP~\cite{shevgoor2015efficiently} learn from history access page offsets or deltas and predict future accesses within a spatial region. Temporal prefetchers like Irregular Stream Buffer (ISB)~\cite{jain2013linearizing} and Domino~\cite{bakhshalipour2018domino} predict temporally correlated memory accesses by recording and replaying the history access sequences. Most rule-based prefetchers require manually configured prefetch degree~\cite{michaud2016best,shevgoor2015efficiently,jain2013linearizing,bakhshalipour2018domino}. Signature Path Prefetcher (SPP)~\cite{kim2016path} uses a path confidence-based lookahead mechanism to balance the prefetching aggressiveness and achieves variable-degree prefetching.

\noindent\textbf{ML-based prefetching.} Several prior works have explored the application of machine learning on data prefetching.
Rahman et al.~\cite{Rahman2015-xk} use logistic regression and decision tree for pattern classification. Hashemi et al.~\cite{mempatterns} present an extensive evaluation of LSTM in learning memory access patterns. Some other works~\cite{peled2018neural, zeng2017long, braun2019understanding} also demonstrate the effectiveness of LSTM in memory access prediction. Srivastava et al. ~\cite{srivastava2019predicting, srivastava2020memmap} use compact LSTM to address the class explosion problem. RAOP~\cite{zhang2020raop} leverages LSTM-based models for virtual address predictions. C-MemMAP~\cite{zhang2021c} combines clustering and meta-models to reduce the model size. Seq2seq modeling~\cite{narayanan2018deepcache} based on LSTM encoder-decoder structure has been applied for memory sequence prediction.
Shi et al.~\cite{shi2021hierarchical} propose Voyager that predicts both page sequence and page offsets using two LSTM models along with a dot-product attention mechanism. Existing ML-based prefetchers use history memory access sequence to predict the next memory access address~\cite{mempatterns,srivastava2020memmap,srivastava2019predicting,zhang2020raop,zhang2021c,shi2021hierarchical,peled2018neural, zeng2017long, braun2019understanding}, which leads to a prefetch degree as one. These models require recurrent greedy/beam search or accepting low-probability candidates to realize higher degree prefetching.

\subsection{Attention}
\label{sec:back_trans}

Transformer~\cite{vaswani2017attention} suggested a sequence model based on multi-head attention mechanism and feed-forward network, dispensing with recurrent structures.


\begin{figure*}[t]
  \centering
  \includegraphics[width=0.75\linewidth]{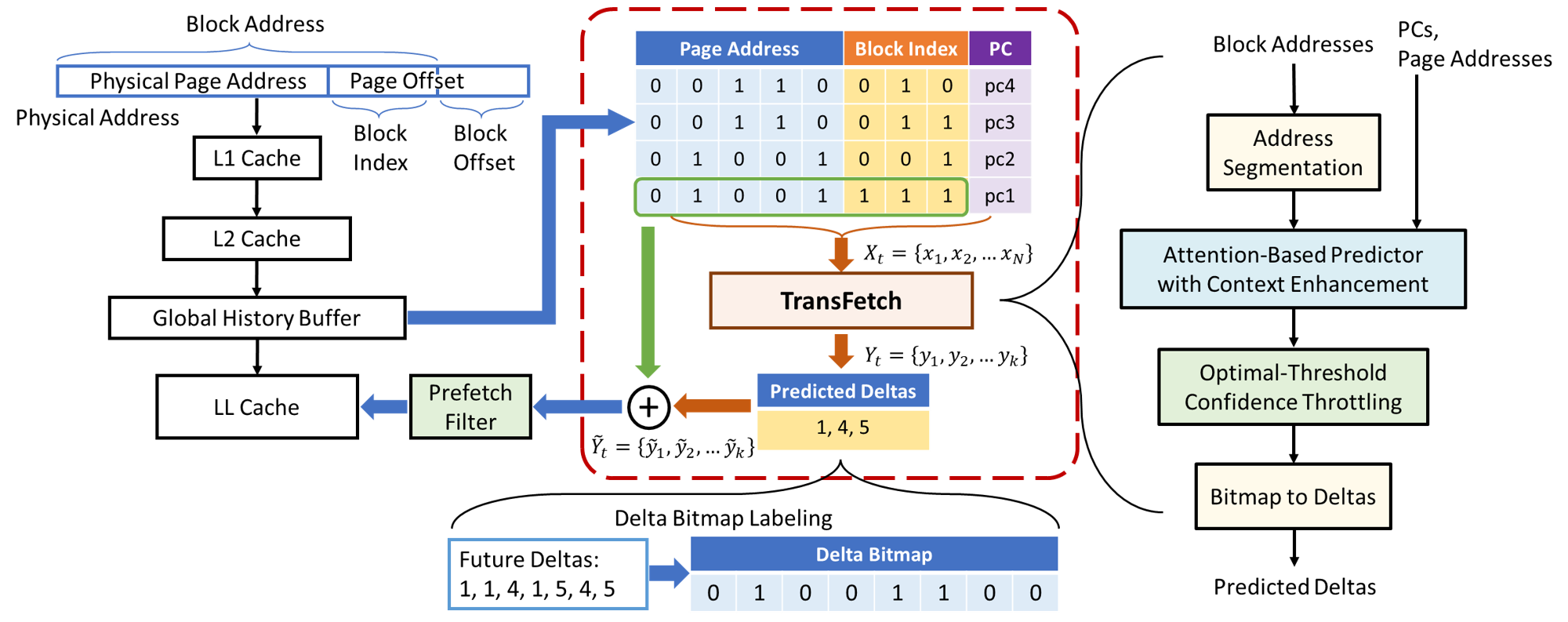}
  \caption{Overall architecture of~\ourwork. We have an input sequence of history memory accesses $X_t = \{x_1,x_2, ..., x_N\}$ and output a set of desired block deltas $Y_t=\{y_1, y_2, ..., y_k\}$ to the current address. The final block address predictions $\widetilde{Y_t}$ are the addition of the current block address and the predicted deltas.}
  \label{fig:overall}
\end{figure*}

\noindent\textbf{Self-attention.} Self-attention takes the embedding of items as input, converts them to three matrices through linear projection, then feeds them into a scaled dot-product attention defined as:
\begin{equation}
\label{eq:att1}
\operatorname{Attention}(Q, K, V)=\operatorname{softmax}\left(\frac{Q K^{T}}{\sqrt{d_{k}}}\right) V
\end{equation}
where $Q$ represents the queries, $K$ the keys, $V$ the values, $d$ the dimension of layer input. 

\noindent\textbf{Multi-head attention.} One self-attention operation can be considered as one "head", we can apply multi-head self-attention (MSA) operation as follows:
\begin{equation}
\begin{aligned}
\label{eq:att2}
\operatorname{MSA}(Q, K, V) &=\operatorname{Concat}\left(\operatorname{head}_{1}, \ldots, \text {head}_{\mathrm{H}}\right) W^{O} \\
\text{head}_{\mathrm{i}} &=\text {Attention}\left(Q W_{i}^{Q}, K W_{i}^{K}, V W_{i}^{V}\right)
\end{aligned}
\end{equation}
where the projection matrics $W_{i}^{Q}, W_{i}^{K}, W_{i}^{V} \in \mathbb{R}^{d \times d}$ and H is the number of heads.

\noindent\textbf{Point-wise feed-forward.} Point-wise feed-forward network (FFN) is defined as follows:
\begin{equation}
\label{eq:att3}
\operatorname{FFN}(x)=\max \left(0, x W_{1}+b_{1}\right) W_{2}+b_{2}
\end{equation}

Previous ML-based prefetchers widely use recurrent neural networks (mainly LSTM)~\cite{mempatterns,srivastava2020memmap,srivastava2019predicting,zhang2020raop,zhang2021c,shi2021hierarchical,peled2018neural, zeng2017long, braun2019understanding}. However, the recurrent structure of RNNs makes the model less practical due to high inference latency. A virtue of attention-based network is high parallelizability. Without recurrent steps, all input positions, hidden representations, and output dimensions can be computed in parallel~\cite{medina2018parallel}. In this work, we will explore using only attention-based networks suggested by~\cite{vaswani2017attention} as a predictor for data prefetching. 

\section{Approach}

In this section we describe~\ourwork, an attention-based prefetcher that uses segmented address as input and achieves variable-degree prefetching through delta bitmap labeling. We formulate the memory access prediction task as a multi-label classification problem and design a neural model to fit the mapping. Since a prefetch must be in the unit of a block, we can consider only the block address space, the address configuration is shown in the left top of Figure~\ref{fig:overall}. 

\noindent\textbf{Problem Formulation.} Let $A_t = \{a_1, a_2, ..., a_N\}$ be the sequence of $N$ history block addresses at time $t$, where $a_t=\{b_t^1, b_t^2,..., b_t^p,..., b_t^{p+c}\}$ represents the block address in binary with $p$-bit page address and $c$-bit block (cache line) index at time $t$. Let $PC_t = \{pc_1, pc_2, ..., pc_N\}$ be the sequence of $N$ history program counters at time $t$. Let $X_t = \{(a_1, pc_1), (a_2, pc_2) ..., (a_N, pc_N)\}$ be the input of the prediction model. Let $Y_t=\{y_1, y_2, ..., y_k\}$ be the set of $k$ outputs associated with the unordered future $k$ block deltas to the current block address. Our goal is to construct meaningful $X_t$ to $Y_t$ that are helpful in data prefetching. The final address predictions $\widetilde{Y_t}=\{\widetilde{y_1}, \widetilde{y_2}, ..., \widetilde{y_k}\}$ are the addition of current block address and the predicted deltas.

\subsection{Overview of~\ourwork}

Figure~\ref{fig:overall} illustrates the overall architecture of~\ourwork~and how the model is applied in a hardware system. 
History block addresses are processed using address segmentation for model inputs, which solves \textit{class explosion} and avoids \textit{tokenization}, as is described in Section~\ref{sec:add_seg}. As a solution for \textit{labeling}, we take future deltas in the form of delta bitmap as training labels. In inference, optimal thresholds for output bitmaps are adapted to adjust the number of outputs (prefetch degree), as in Section~\ref{sec:label}.
To reduce inference \textit{latency}, a powerful and parallelizable attention-based network is designed for learning the mapping between input and output, as is described in Section~\ref{sec:model}. To further offset the latency, a distance prefetching scheme is discussed in Section~\ref{sec:distance}.

\begin{figure*}[t]
\centering
\subfloat[Address segmentation]
{\includegraphics[width=0.2\linewidth]{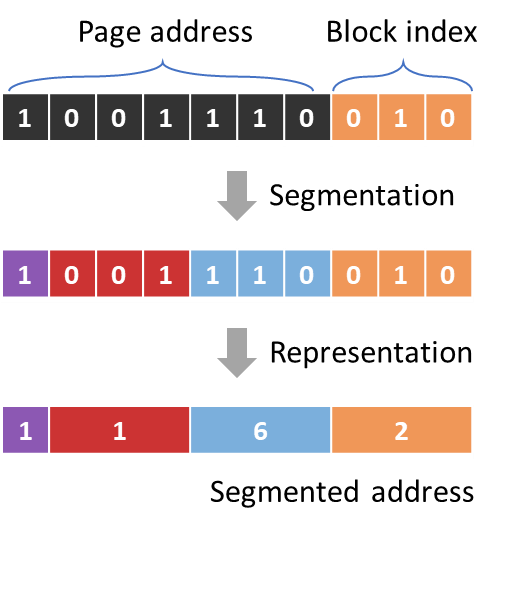}\label{fig:seg1}}
\subfloat[Examples of memory access patterns visualized in segmented addresses.]
{\includegraphics[width=0.68\linewidth]{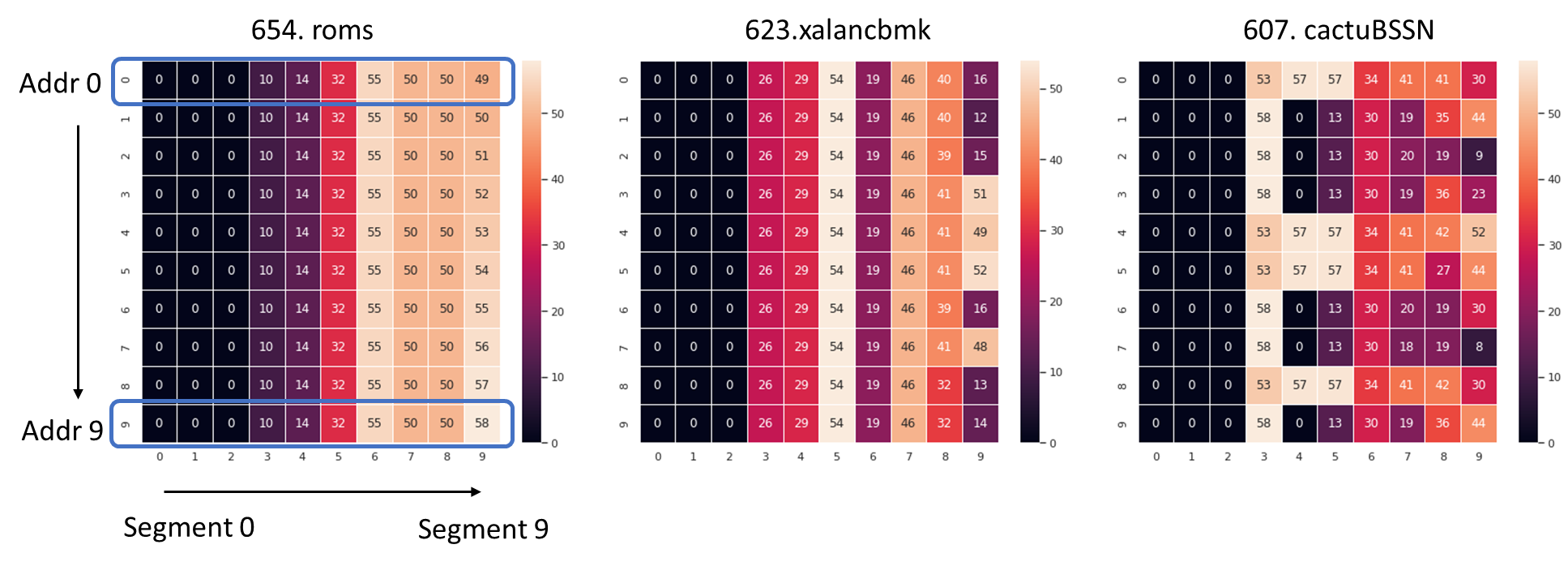}\label{fig:seg2}}
\caption{Address segmentation approach and memory access pattern cases visualized under segmented addresses.}
\label{fig:seg}
\end{figure*}

\subsection{Address Segmentation}
\label{sec:add_seg}
We propose a simple approach called address segmentation (AS) to solve the class explosion problem in memory access prediction, keeping all the information in an address and avoiding tokenization. 

Considering a block address with $p$-bit page address and $c$-bit block index, we can split this block address to $S=\lceil{\frac{p+c}{s}}\rceil$ segments, each with $s$ bits. In this way, each segment can be represented in an integer within $[0 - 2^s)$. This range can be tuned appropriately for direct model input. One address then can be represented as a vector in dimension $S$. 

There are two special cases. The first is when $s=1$, the model input is a binary of an address. 
This case reveals the detail of an address in the highest granularity but requires a model to learn the correlation of each bit. The other case is when $s=c$, which means using the block size $c$ as the segmentation basis and split the block address to $S=\lceil{\frac{p}{c}}\rceil+1$ segments, the segment vocabulary is $2^c$, 64 for $c=6$. This case keeps the feature of internal page patterns and reduces the input dimension compared with binary inputs. 

Figure~\ref{fig:seg1} illustrates address segmentation for case $s=c$. Figure~\ref{fig:seg2} visualizes three example pieces of memory access sequences from SPEC CPU 2017~\cite{SPEC2017} in form of this case, which illustrates the advantages of AS for preprocessing. While in application \textit{654.roms}, the pattern lies mainly within a page (the column of segment 9), \textit{623.xalancbmk} shows memory access skipping beyond the page limit. Furthermore, \textit{607.cactuBSSN} shows more complex patterns, e.g., skippings between pages. AS keeps the information of an absolute address compared with solely delta, offset, or page inputs. In addition, AS avoids token dictionaries, saves storage space, and can process unknown input classes.

\subsection{Variable-Degree Prefetching}
\label{sec:label}
\subsubsection{Delta Bitmap Labeling}
We use delta bitmaps as the format of labels and outputs. We aim to predict multiple unordered future deltas to the current block, as is shown in the bottom part of Figure~\ref{fig:overall}.
The labels are acquired from offline traces for training. First, future deltas $\mathbf{y_d}$ are collected within a look forward window $W$.
Then, a delta bitmap at size $B$ is filled to label the appearance of deltas by an arbitrary mapping rule $f: \mathbf{y_d} \rightarrow \mathbf{y_b}$, where $\mathbf{y_b}$ is the labeled bitmap, which can be used for multi-label model training. By designing the delta bitmap size $B$ to be larger than a page, our model can learn and predict inter-page patterns, which addresses the weakness rule-based spatial prefetchers like BOP~\cite{michaud2016best} and VLDP~\cite{shevgoor2015efficiently}.

\subsubsection{Optimal-Threshold Confidence Throttling}
A neural network can be designed to output the probability of each bit being positive in a bitmap. We define this probability as \textit{prefetch confidence} for the corresponding deltas in bitmap. 
Instead of using a fixed threshold, e.g. 0.5, to binarize the model output, we find the optimal threshold that maximizes the F1-score~\cite{saito2015precision} in the step of model validation, between model training and testing. In inference, the output vector of prefetch confidence can be binarized using the optimal threshold, which forms the output delta bitmap.
Then, the inverse mapping $f^{-1}$ converts the output delta bitmap to the predicted delta vectors. 

Using delta bitmap and optimal-threshold confidence throttling together for inference, our approach enables variable-degree prefetching: a model can predict and request prefetching variable number of future cache lines in one inference step.

\subsection{Attention-Based Predictor}
\label{sec:model}
\begin{figure}[t]
  \centering
  \includegraphics[width=0.92\linewidth]{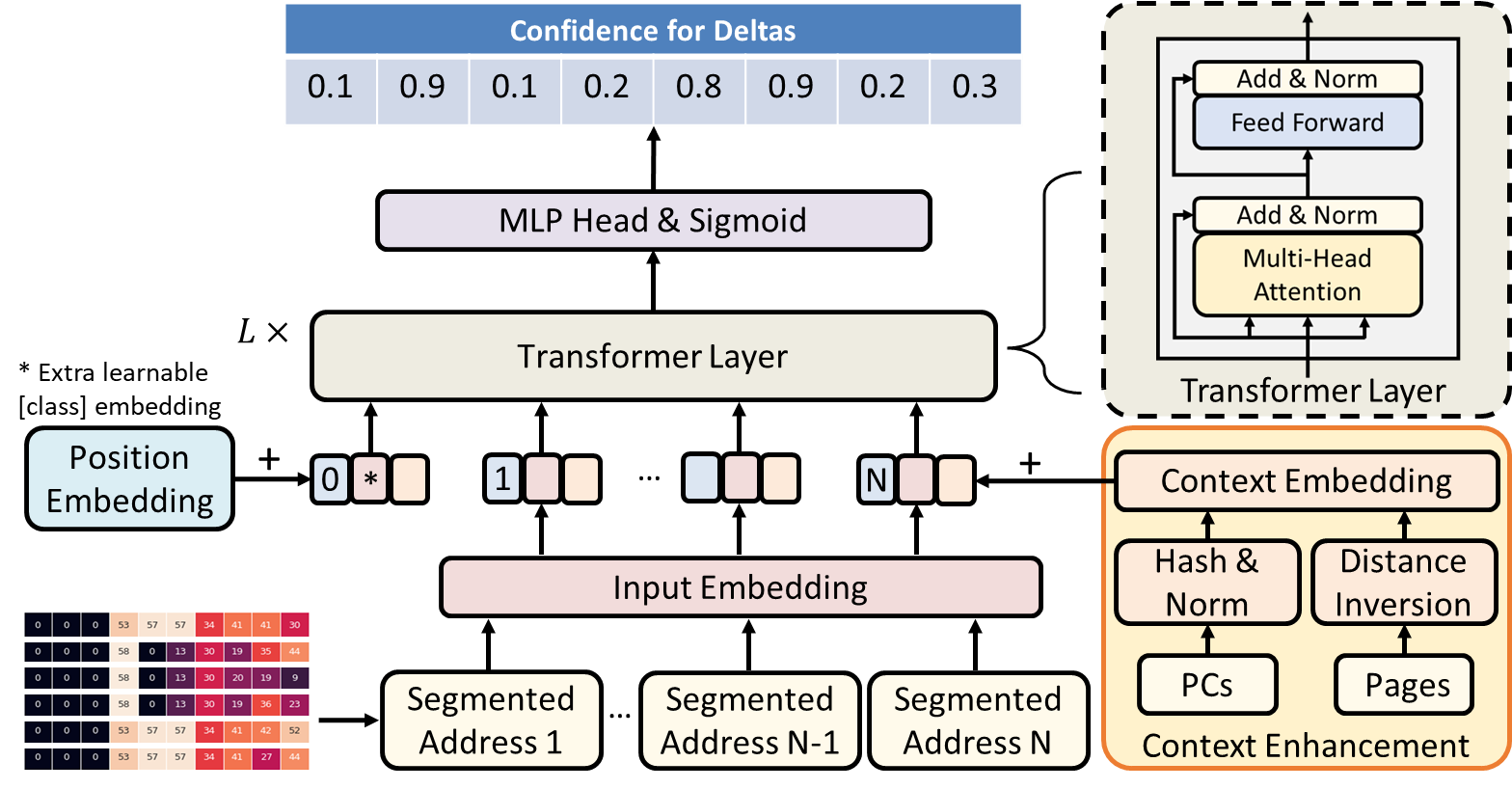}
  \caption{Attention-based memory access predictor with context enhancement.}
  \label{fig:trans_cx}
\end{figure}

With the above well-defined input and output, we design an attention-based network to learn the mapping from the input segmented addresses to the delta bitmap labels. The model structure is depicted in Figure~\ref{fig:trans_cx}. First, we describe the basic model. Then we introduce context enhancement that utilizes more context information to boost the model performance. 

\subsubsection{Basic Model}
\label{sec:model_basic}

Our model input is a 2D sequence of segmented addresses: $\mathbf{a}_{S}=[\mathbf{a}_{S}^1;\mathbf{a}_{S}^2;...;\mathbf{a}_{S}^N] \in \mathbb{R}^{N \times S}$  where $N$ is the number of history addresses and $S$ is the dimension of a segmented address. 

We flatten the input sequence and map to $D$ dimensions using an input embedding layer, where $D$ is the hidden dimension of Transformer layers. Inspired by BERT~\cite{devlin2018bert} and ViT ~\cite{dosovitskiy2020image}, a trainable "classification token" denoted as $\mathbf{x}_{cls}$ is prepended to the input sequence, whose state is a comprehensive representation of the input sequence. 
We apply learnable 1D position embeddings~\cite{dosovitskiy2020image} to incorporate temporal information to input vector.
The addition of input embeddings and position embeddings are fed into a the Transformer layers, which can be expressed as:
\begin{equation}
    \mathbf{z}_{0} =\left[\mathbf{x}_{cls } ; \mathbf{a}_{S}^{1} \mathbf{E} ; \mathbf{a}_{S}^{2} \mathbf{E} ; \cdots ; \mathbf{a}_{S}^{N} \mathbf{E}\right]+\mathbf{E}_{\text {pos }}
\end{equation}
where $\mathbf{z}_{0}$ represents input sequence to the Transformer layers, $\mathbf{E}$ is the input embedding and $\mathbf{E}_{\text{pos}}$ is the position embedding.

The Transformer layer is based on multi-head attention and feed-forward network as described in Section~\ref{sec:back_trans}. The output of $L$ stacked Transformer layers will be fed into a multi-layer perceptron (MLP) head for multi-label classification, which is the same dimension as the delta bitmap size $B$. A sigmoid activation is applied to each output dimension and outputs the probability of this dimension being positive, which we use as the prefetch confidence for deltas.

\subsubsection{Context Enhancement} 
The model can incorporate richer input features to boost the model performance using the same method as position embedding. 

First, we incorporate program counters (PC), which are commonly used to help detecting memory patterns~\cite{somogyi2006spatial,jain2013linearizing}. To avoid tokenization, we use a $HASH\_BITS$ bit length folding method~\cite{maurer1975hash} as the hash function to compress the PC value $pc_n$. The hashed value is divided by $2^{HASH\_BITS}$ for normalization, the processed PC input vector is $\mathbf{c_{pc}}=[c_{pc}^1;c_{pc}^2;...;c_{pc}^N]$, $c_{pc}^n$ is defined as:
\begin{equation}
c_{pc}^n =  \frac{hash(pc_n)}{2^{HASH\_BITS}}
\end{equation}

Second, we incorporate page distance (PD) based on the hypothesis that adding weights to input addresses through the inversion of page distances can improve the model prediction performance:

\begin{equation}
c_{pd}^n =  \frac{1}{|page_n-page_1|+1}
\end{equation}
where $page_n$ is the page address at history $n$, $page_1$ is the current page address, the PD input vector is $\mathbf{c_{pd}}=[c_{pd}^1;c_{pd}^2;...;c_{pd}^N]$.


The context input is the concatenation of $\mathbf{c_{pc}}$ and $\mathbf{c_{pd}}$.
A linear projection $\mathbf{E}_{\text {ce}}$ is applied for context embedding that maps the context input vector to the same dimension of the Transformer input. Therefore, the input embedding, position embeddings and the context embedding can be added as described in Equation~\ref{eq:ce}. 
\begin{equation}
    \label{eq:ce}
    \mathbf{z}_{0}' = \mathbf{z}_{0} + \mathbf{[c_{pc};c_{pd}]} \mathbf{E}_{\text {ce}}
\end{equation}
\subsubsection{Loss Function} For our multi-label classification problem, we use binary cross-entropy loss defined as below:
\begin{equation}
\mathcal{L}=-\frac{1}{N} \sum_{i=1}^{N} y_{i} \log \left(p\left(y_{i}\right)\right)+\left(1-y_{i}\right) \log \left(1-p\left(y_{i}\right)\right)
\end{equation}
where $y_i$ is the label and $p(y_i)$ is the predicted probability for sample $i$ being $True$. For multi-label training, each dimension is considered independent and the loss is summed.


\subsection{Distance Prefetching}
\label{sec:distance}
A real hardware implementation will incur some latency. Attention-based model is feasible for high parallel implementations. According to Equation~\ref{eq:att1} - \ref{eq:att3} and Figure~\ref{fig:trans_cx}, the network inference latency under a fully paralleled implementation can be estimated as: 
\begin{equation}
\label{eq:latency}
\begin{array}{l}
 Latency=\underbrace{T_{mm_e}+T_{add}}_{\text{Embeddings}}+\underbrace{T_{mm_{h}}+T_{av}}_{\text{Output head}}+
 \\ L\times[\underbrace{\underbrace{4 T_{mm_a}+3T_{av}}_{\text{Multi-head attention}}+T_{mm_{f}}+2(T_{add}+T_{norm})]}_{\text{Transformer layer}}
\end{array}
\end{equation}
where $T_{mm_e}$ is the latency of matrix multiplication for the input embeddings, $T_{add}$ is the vector addition latency, $T_{mm_{h}}$ is the MLP head latency, $T_{av}$ is the latency for activation functions, mask, and scale operations, $T_{mm_a}$ is the latency of multi-head attention, $T_{norm}$ is the normalization latency. $L$ is the number of Transformer layers.

While attention-based network reduces inference latency, we can further offset the latency by skipping the inference slot and predict the future memory accesses in a distance. The model for distance prefetching can by easily trained through distance labeling, i.e., collecting labels by skipping the estimated inference latency.

\section{Experiments}

\subsection{Benchmarks}
We evaluate~\ourwork~and the baselines using the application traces generated from benchmarks \emph{SPEC CPU 2006}~\cite{jaleel2010memory}, \emph{SPEC CPU 2017}~\cite{SPEC2017}, and \emph{GAP}~\cite{beamer2015gap} using SimPoint ~\cite{perelman2003using}. 
After skipping 1M instructions for warm-up, we use 100M instructions for experiments.
We use the first 40M instructions for model training, the next 10M instructions for validation, tuning, and generating optimal thresholds, and the last 50M instructions for evaluation\footnote{The code is available at: \url{https://github.com/pgroupATusc/TransFetch.git}}.
\begin{table}[h]
  \caption{Benchmark statistics}
  \label{tab:dataset}
  \begin{center}
    \begin{small}
  \begin{tabular}{lccccc}
    \toprule
    BMKs &\# PCs &\# Addresses &\# Pages\ &\# Deltas\\
    \midrule
    SPEC 06 & 23$\sim$893 &60.0K$\sim$2.21M & 2.51K$\sim$88.9K & 23.6K$\sim$2.01M\\
    SPEC 17 &26$\sim$1126 &62.1K$\sim$1.78M & 7.99K$\sim$ 0.26M& 3.18K$\sim$0.72M\\
    GAP &63$\sim$118 &0.56M$\sim$1.25M & 8.27K$\sim$ 27.2K & 0.30M$\sim$1.20M\\
  \bottomrule
\end{tabular}
\end{small}
\end{center}
\end{table}

Table~\ref{tab:dataset} shows the number of unique program counters (PCs), addresses, page addresses, and deltas. If using tokenization, a token dictionary needs to store the mapping of the unique values to tokens, which is consumes storage. Our method discards tokenization and saves up to table of length 2.01M compared with delta inputs, and up to table of length 0.26M compared with page \& offset inputs, given the same level of model complexity.

\subsection{Prediction Evaluation}
We evaluate the prediction performance of the model by comparing the predicted deltas to the labels. With fixed labels, we vary the model backbones (feature extractor) and inputs to understand the advantages of~\ourwork.

\begin{table}[h]
  \caption{Model configuration}
  \label{tab:model}
  \begin{center}
    \begin{small}
  \begin{tabular}{llclc}
    \toprule
    &Configuration&Value&Configuration&Value\\
    \midrule
    Input/output & Delta bound & $\pm$128 &Delta bitmap $B$ & 256\\
    & History $N$ & 9 & Look-forward $W$ & 128\\
     \midrule
    Attention & Dimension $D$ & 128 &  MLP head layer & 1\\
    &Head number & 4 & Layer $L$ & 2\\
  \bottomrule
\end{tabular}
\end{small}
\end{center}
\end{table}

\begin{table*}[t]
\caption{Comparison of model backbones and input methods}
\label{tab:backbones}
\begin{center}
\begin{small}
\begin{tabular}{ccccccccccccc}
\toprule
    & & &\multicolumn{3}{c}{SPEC 06} & \multicolumn{3}{c}{SPEC 17}& \multicolumn{3}{c}{GAP}\\
    \cmidrule(lr){4-6} \cmidrule(lr){7-9}\cmidrule(lr){10-12}
    Backbone & Input & \# Params (K) &Precision&Recall&F1-score &Precision&Recall&F1-score &Precision&Recall&F1-score\\
    \midrule
         & Delta$^\ast$  & 477 & 0.5877 &0.4218 &0.4911 &0.6017 &0.4215 &0.4957 &0.5618 &0.0670 &0.1197\\
        & Page \& offset$^\ast$ &625&0.6323 &0.4356 &0.5158 &0.6208 &0.4257 &0.5051 &0.5994 &0.1372 &0.2233\\
        & 1-bit AS &812 &\textbf{0.7158}	&\textbf{0.5039}	&\textbf{0.5914}	&\textbf{0.6763}	&\textbf{0.4519}	&\textbf{0.5418}	&0.5835	&\textbf{0.3594	}&\textbf{0.4448}\\
    LSTM    & 4-bit AS &451 &0.6122	&0.4712	&0.5325	&0.6177	&0.4364	&0.5115	&0.4281	&0.3361	&0.3765\\
    (256,1,256)  & 6-bit AS$^\dagger$ &394 &0.6291 &0.4520	&0.5260	&0.6172	&0.4383	&0.5126	&\textbf{0.6096}	&0.3403	&0.4368\\
        & 8-bit AS &394 & 0.6513	&0.4938	&0.5617 &0.6160	&0.4457	&0.5172 & 0.5762 &0.3311 &0.4206\\
        & 12-bit AS &369 & 0.6110	&0.4510	&0.5190 &0.6072	&0.4397	&0.5101 &0.5106	&0.3467	&0.4130\\
        & 16-bit AS &361 & 0.6022	&0.4147	&0.4911 &0.6519	&0.4407	&0.5259 &0.5406	&0.3393	&0.4169\\
    \midrule
    & Delta$^\ast$  & 889 & 0.5633	&0.4179	&0.4798	&0.5833	&0.4101	&0.4816	&0.3618	&0.1350	&0.1966\\
        & Page \& offset$^\ast$ & 1679 &0.6871	&0.4921	&0.5735	&0.6110	&0.4433	&0.5138	&0.5219	&0.2024	&0.2917\\
        & 1-bit AS & 8153&0.6992 &\textbf{0.5298}	&\textbf{0.6028}	&0.6231	&0.4833	&0.5444	&0.5639	&0.3451	&0.4282\\
    TCN    & 4-bit AS & 436& 0.7027	&0.5173	&0.5959 &\textbf{0.6679}	&0.4623 &\textbf{0.5464} &0.4522	&0.3530	&0.3965\\
    ($l_{in}$,1,4,256)    & 6-bit AS$^\dagger$ & 201&\textbf{0.7135}	&0.5134	&0.5971	&0.6169	&\textbf{0.4892}	&0.5457	&\textbf{0.5877}	&\textbf{0.3551}	&\textbf{0.4358}\\    
        & 8-bit AS & 132 & 0.6033	&0.4138	&0.4909	&0.6265	&0.4335	&0.5124	&0.4425	&0.3223	&0.3730\\
        & 12-bit AS &43  &0.6005	&0.4127	&0.4892	&0.5989	&0.4328	&0.5024	&0.3976	&0.3213	&0.3554\\
        & 16-bit AS &29  &0.5908	&0.4079	&0.4826	&0.5972	&0.4316	&0.5011	&0.3966	&0.3250	&0.3572\\
    \midrule
     & Delta$^\ast$  & 431& 0.5112	&0.4338	&0.4693	&0.5623	&0.3947	&0.4638	&0.1956	&0.1986	&0.1971\\ 
        & Page \& offset$^\ast$ & 431 &0.5989	&0.5637	&0.5808	&0.5699	&0.5123	&0.5396	&0.2676	&0.4012	&0.3211\\
        & 1-bit AS &433 &0.6856 &0.7439 &0.7136 &\textbf{0.6025} &0.6484 &0.6246 & 0.4020 &0.8418 &0.5442\\
    Attention    & 4-bit AS &433 &0.6637	&0.7337	&0.6969	&0.6020	&0.6118	&0.6069	&0.3876	&0.7337	&0.5072\\
    (128,4,2,256)    & 6-bit AS$^\dagger$ &432 &	\textbf{0.6997}	&\textbf{0.7565}	&\textbf{0.7270}	&0.6018	&\textbf{0.6901}	&\textbf{0.6429} &\textbf{0.4170}	&\textbf{0.8429}	&\textbf{0.5580}\\
        & 8-bit AS &432 & 0.6869	&0.6918	&0.6893	&0.5331	&0.5701	&0.5510	&0.3174	&0.7068	&0.4381\\
        & 12-bit AS &432 & 0.6799	&0.6865	&0.6832	&0.5796	&0.5443	&0.5614	&0.3431	&0.7162	&0.4639\\
        & 16-bit AS &432 & 0.6693	&0.6935	&0.6812	&0.5424	&0.5215	&0.5317	&0.3097	&0.7020	&0.4298\\
    \bottomrule
    \multicolumn{12}{l}{\small $\ast$ Delta input and page \& offset input need tokenization and require extra storage for token dictionaries.}\\
    \multicolumn{12}{l}{\small $\dagger$ 6-bit AS is when the segment size equals the block index size, i.e. $s=c$.} \\
\end{tabular}
\end{small}
\end{center}
\label{tab:ablation}
\end{table*}

\subsubsection{Implementation}
The configuration of~\ourwork~is shown in Table~\ref{tab:model}. The models are trained using ADAM~\cite{kingma2014adam} optimizer with decayed learning rate. 
We set the delta bound as $\pm 128$ that can skip the page boundary of 64, which determines the bitmap size as 256. 

\begin{table*}[h]
\caption{Ablation study of program counter (PC) and page distance (PD)}
\label{tab:ablation}
\begin{center}
\begin{small}
\begin{tabular}{cccccccccccc}
\toprule
    &&&\multicolumn{3}{c}{SPEC 06} & \multicolumn{3}{c}{SPEC 17}& \multicolumn{3}{c}{GAP}\\
    \cmidrule(lr){4-6} \cmidrule(lr){7-9}\cmidrule(lr){10-12}
    Method &PC& PD &Precision&Recall&F1-score &Precision&Recall&F1-score &Precision&Recall&F1-score \\
    \midrule
    Basic  & & &{0.6997}	&{0.7565}	&{0.7270}	&0.6018	&{0.6901}	&{0.6429} &{0.4170}	&\textbf{0.8429} &0.5580\\
    +$\mathbf{c_{pc}}\mathbf{E}_{\text {ce}}$ &\checkmark & & 0.7430	&\textbf{0.7777}	&0.7599 &0.8080	&0.5314	&0.6412 & 0.6245 &0.6268 &0.6256\\
    +$\mathbf{c_{pd}}\mathbf{E}_{\text {ce}}$ & &\checkmark  & 0.8493	&0.7117	&0.7744 &0.7556	&0.5881	&0.6614 &0.5856	&0.6377	&0.6105\\
    +$\mathbf{[c_{pc};c_{pd}]}\mathbf{E}_{\text {ce}}$ &\checkmark &\checkmark  &\textbf{0.8638}	&0.7217	&\textbf{0.7864} &\textbf{0.8144}	&\textbf{0.6496}	&\textbf{0.6735} &\textbf{0.6634}	&0.6344 &\textbf{0.6486}\\
    \bottomrule
\end{tabular}
\end{small}
\end{center}
\label{tab:ablation}
\end{table*}

\subsubsection{Backbones}
To evaluate the contribution of attention layers, under the same input and output configuration in Table~\ref{tab:model}, we implement three neural networks as the backbones of our framework:
\begin{itemize}
    \item \textbf{LSTM}~\cite{hochreiter1997long} with hidden dimension = 256, number of layers = 1, and output dimension = 256, indicated as (256, 1, 256).
    \item \textbf{TCN}~\cite{lea2017temporal} with hidden dimension same as input sequence length $l_{in}$, channel = 1, filter size = 4, and output dimension = 256, indicated as ($l_{in}$, 1, 4, 256).
    \item \textbf{Attention}~\cite{vaswani2017attention} as in Table~\ref{tab:model}, with hidden dimension = 128, number of heads = 4, depth = 2, and output dimension = 256, indicated as (128, 4, 2, 256).
\end{itemize}

\subsubsection{Inputs}
To evaluate the contribution of address segmentation, we implement three input methods:
\begin{itemize}
    \item \textbf{Delta} input uses the jumps between consecutive memory access addresses. A value-to-token dictionary is required; this requires extra storage space.
    \item \textbf{Page \& offset} splits an address only to page address and page offset. The page addresses also need tokenization and use extra storage for token dictionary. The offsets can be directly fed into the model.
    \item \textbf{Address segmentation (AS)} splits an address to segments and avoids tokenization. Particularly, when the segmentation bit = 6, the segment size is same as the block index.
\end{itemize}

\subsubsection{Metrics}
We use precision, recall, and F1-score~\cite{powers2020evaluation} to evaluate the memory access prediction performance of the models.

\subsubsection{Output threshold}
For all the implemented models, we determine the optimal threshold through a grid search to achieve the highest F1-score in validation.

\subsubsection{Results}
Table~\ref{tab:backbones} shows the prediction performance of the implemented backbones under various inputs. For a fair comparison, the models are tuned with the same order of complexity, except the TCN whose model size is influenced by the input format. For each backbone, the trend is clear that 1-bit AS and 6-bit AS as inputs result in higher performance than delta input, page \& offset inputs, and other AS methods. Specifically, LSTM is stronger in bit-wise input while attention shows higher performance in 6-bit segmentation. Address segmentation achieves 0.26 - 0.36 higher F1-score than delta inputs and 0.15 - 0.24 higher F1-score than page \& offset inputs. Comparing among different backbones, we observe that attention-based models typically acquire higher recall than LSTM and TCN, which leads to 0.10 - 0.13 higher F1-score.

\subsubsection{Effectiveness of Context Enhancement}
To evaluate the influence of context enhancement (CE), we conduct ablation studies on program counter (PC) and page distance (PD). 
When we introduce both PC and PD, the enhanced model achieves the highest precision and F1-score for all the three benchmarks. The F1-score improves by 3.1\% - 9.1\% when context enhancement is introduced.

\subsection{Prefetching Evaluation}
We evaluate the prefetching performance of~\ourwork~by comparing with the state-of-the-art rule-based prefetchers and ML-based prefetchers. Both the input methods and the labeling methods can be different among these prefetchers. We follow their original designs and compare the overall contributions to the improvement of IPC under the same data set and simulation environment.

\subsubsection{Simulator}
We evaluate our approach using the simulation framework released by the 2021 ML-Based Data Prefetching Competition, which is based on ChampSim~\cite{ChampSim}. 
The simulator parameters are shown in Table~\ref{tab:sim_param}. 
We simulate all prefetchers at the last-level cache (LLC). There is no prefetching for other cache levels. 
\subsubsection{Baseline Prefetchers}
We compare~\ourwork~with state-of-the-art rule-based prefetchers and ML-based prefetchers:
\begin{itemize}
    \item \textbf{Rule-based prefetchers} including spatial prefetchers BOP~\cite{michaud2016best}, VLDP~\cite{shevgoor2015efficiently}, and SPP~\cite{kim2016path} with variable degree; temporal prefetchers ISB~\cite{jain2013linearizing} and Domino~\cite{bakhshalipour2018domino}.
    \item \textbf{ML-based prefetchers} including Embedding-LSTM~\cite{hashemi2018learning} with delta inputs, Clustering-LSTM~\cite{hashemi2018learning} with address space clustering, and Voyager~\cite{shi2021hierarchical}.
\end{itemize}

\begin{table}[t]
  \caption{Simulation parameters}
  \label{tab:sim_param}
  \begin{center}
    \begin{small}
  \begin{tabular}{lc}
    \toprule
    Parameter&Value\\
    \midrule
    CPU & 4 GHz, 4-wide OoO, 256-entry ROB, 64-entry LSQ\\
    L1 I-cache & 64 KB, 8-way, 8-entry MSHR, 4-cycle\\
    L1 D-cache & 64 KB, 12-way, 16-entry MSHR, 5-cycle\\
    L2 Cache&	1 MB, 8-way, 32-entry MSHR, 10-cycle\\
    LL Cache&	8 MB, 16-way, 64-entry MSHR, 20-cycle\\
    DRAM &	 $t_{RP}=t_{RCD}=t_{CAS} =$  12.5 ns, 2 channels,\\ 
& 8 ranks, 8 banks, 32K rows, 8GB/s bandwidth per core\\
  \bottomrule
\end{tabular}
\end{small}
\end{center}
\end{table}




\subsubsection{Metrics}
We use prefetch accuracy, coverage, and IPC improvement~\cite{srinivasan2004prefetch} to evaluate the prefetching performance. Defining a useful prefetch as the prefetched line being referenced by the application before it is replaced, we have:
\begin{itemize}
    \item \textbf{Prefetch accuracy} as the ratio of useful prefetches to the overall prefetches;
    \item \textbf{Prefetch coverage} as the ratio of useful prefetches to the overall cache misses;
    \item \textbf{IPC improvement} as the percentage increase of instructions per cycle.
\end{itemize}




\begin{figure}[h]
\centering
\subfloat[Optimal threshold]
{\includegraphics[width=0.45\linewidth]{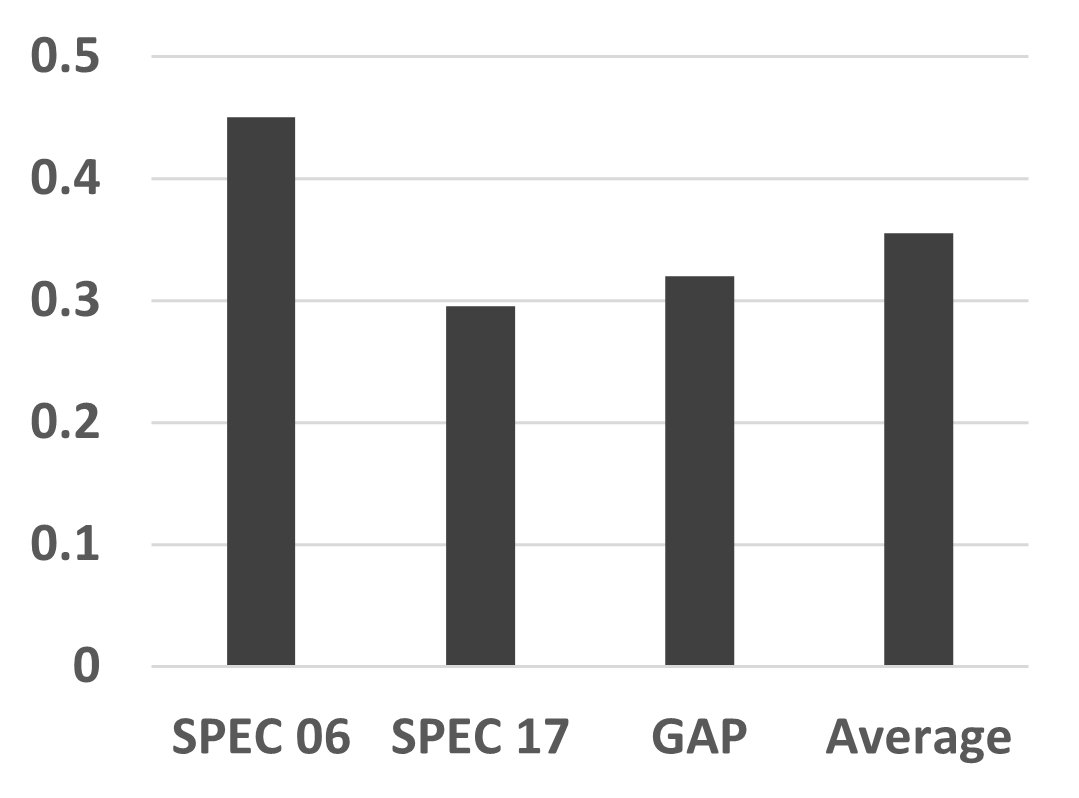}\label{fig:deg1}}
\subfloat[Mean prefetch degree]
{\includegraphics[width=0.45\linewidth]{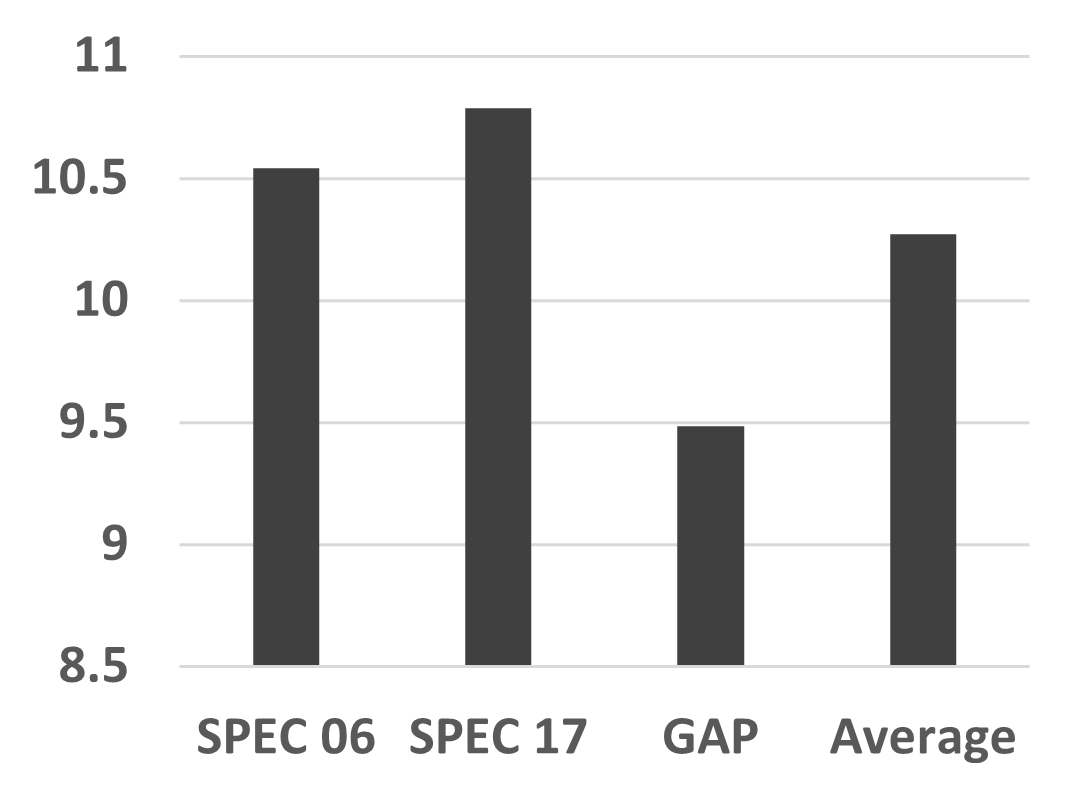}\label{fig:deg2}}
\caption{Optimal threshold and mean prefetch degree.}
\label{fig:deg}
\end{figure}


\subsubsection{Prefetch Degrees}

Figure~\ref{fig:deg1} shows the adapted optimal thresholds for~\ourwork. The average thresholds are 0.448, 0.296, and 0.323 for benchmarks SPEC 06, SPEC 17, and GAP, respectively. The overall average threshold is 0.355. Figure~\ref{fig:deg2} shows the mean prefetch degrees after throttled from optimal thresholds. The overall average degree is 10.271 across all applications. According to the degree results, we set the overall degree of rule-based baseline prefetchers as 10 for a fair comparison. For ML-based prefetchers, we prefetch predictions with the 10 top probabilities.

\subsubsection{Simulation Results}

\begin{figure}[t]
  \centering
  \includegraphics[width=\linewidth]{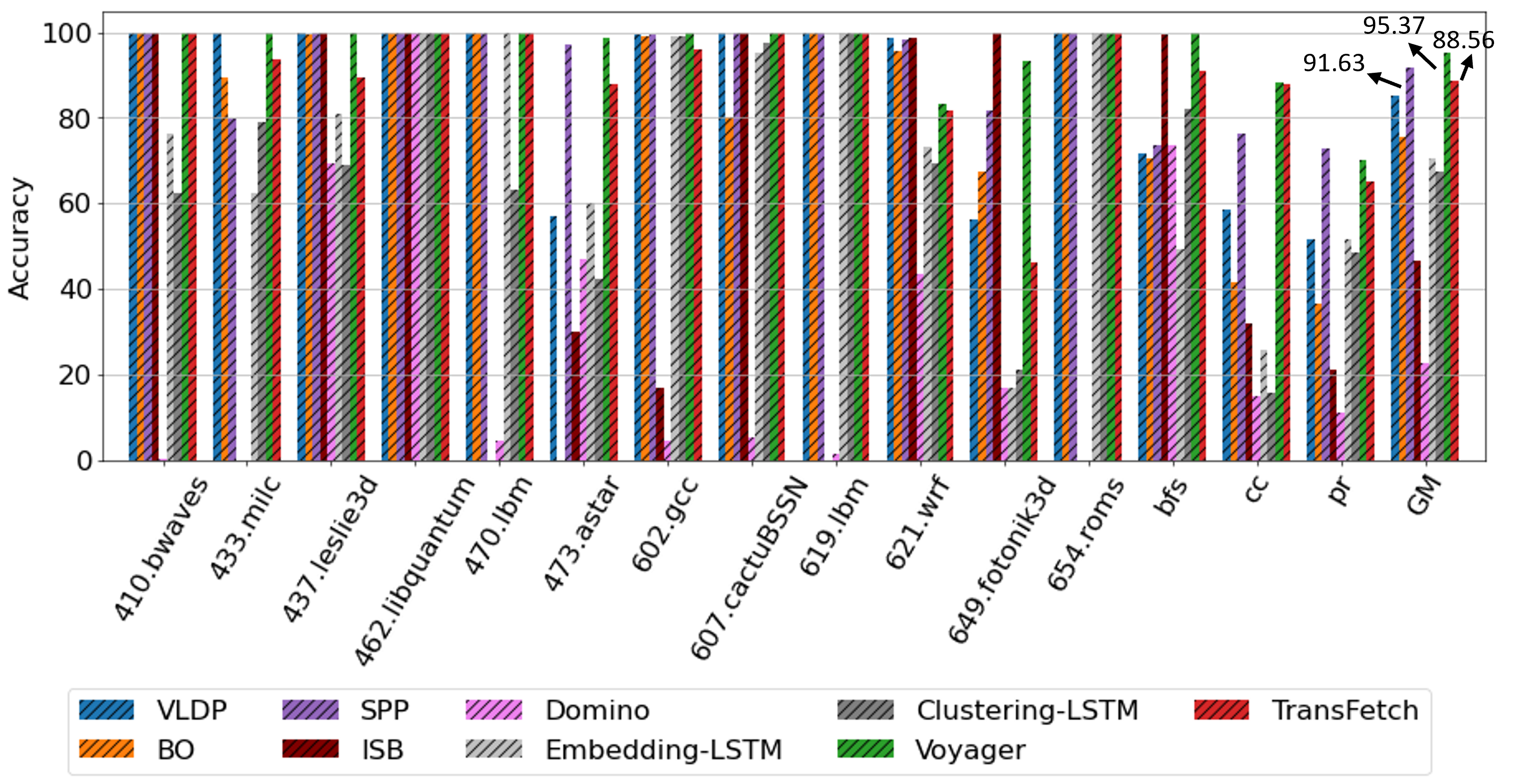}
  \caption{Prefetch accuracy of~\ourwork~and baselines.}
  \label{fig:acc}
\end{figure}

Figure~\ref{fig:acc} illustrates the prefetch accuracy.
Voyager achieves the highest average accuracy at 95.37\% among all prefetchers. SPP achieves accuracy at 91.63\%, the highest among rule-based prefetchers. 
~\ourwork~has accuracy at 88.56\% that is the third highest among all prefetchers. VLDP, BOP, ISB, Domino, Embedding-LSTM, and Clustering-LSTM achieve lower accuracy at 85.08\%, 75.69\%, 46.49\%, 22.70\%, 70.79\%, and 67.54\%, respectively.

\begin{figure}[t]
  \centering
  \includegraphics[width=\linewidth]{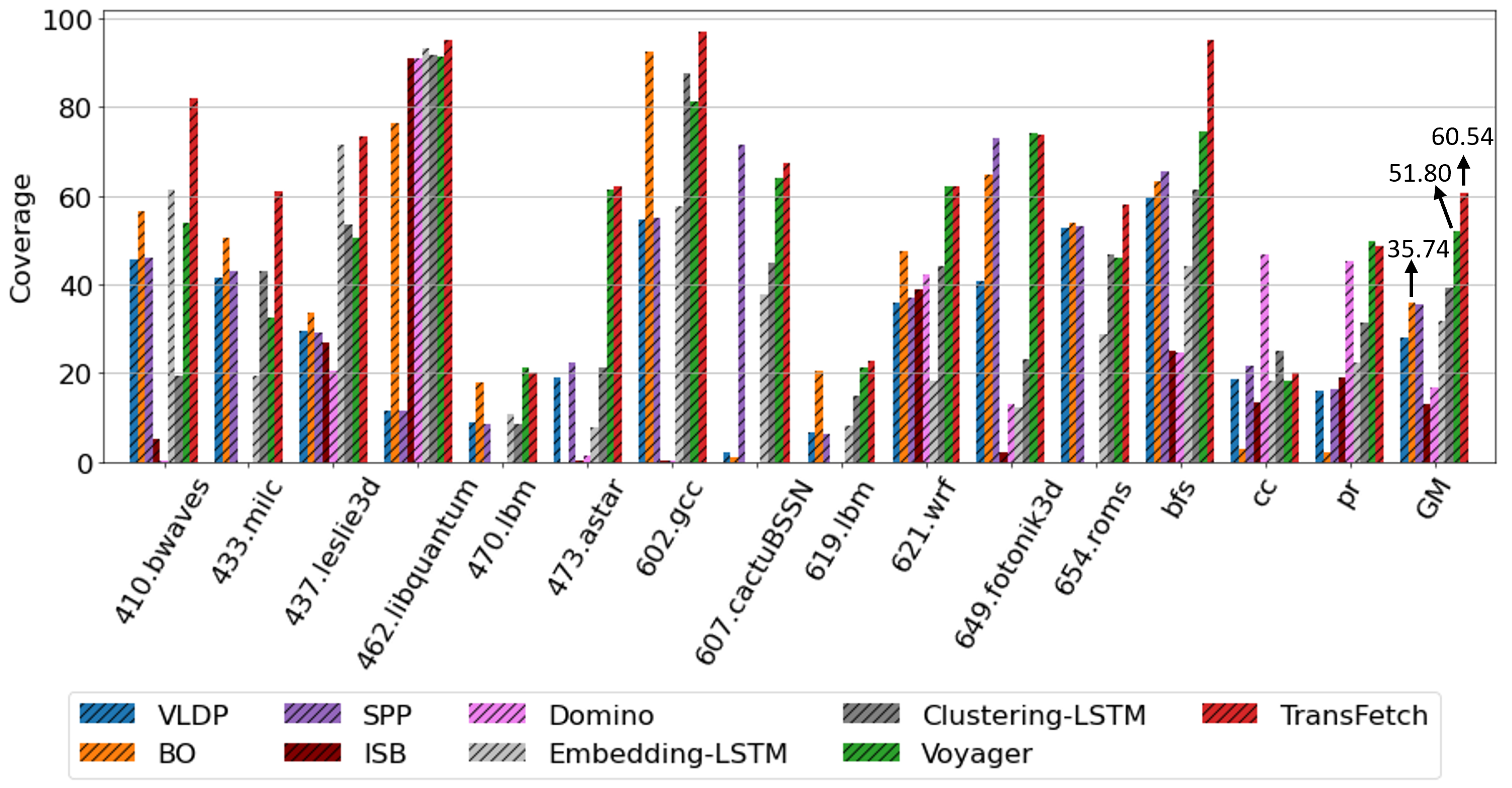}
  \caption{Prefetch coverage of~\ourwork~and baselines.}
  \label{fig:cov}
\end{figure}

Figure~\ref{fig:cov} shows the prefetch coverage. In an average,~\ourwork~achieves the highest coverage at 60.54\%, compared with Voyager at 51.80\%. BOP achieves 35.74\%, which is the highest among rule-based prefetchers. VLDP, SPP, ISB, Domino, Embedding-LSTM, and Clustering-LSTM achieve lower coverage at 28.09\%, 35.47\%, 12.80\%, 16.57\%, 31.86\%, and 39.16\%, respectively.

\begin{figure}[t]
  \centering
  \includegraphics[width=\linewidth]{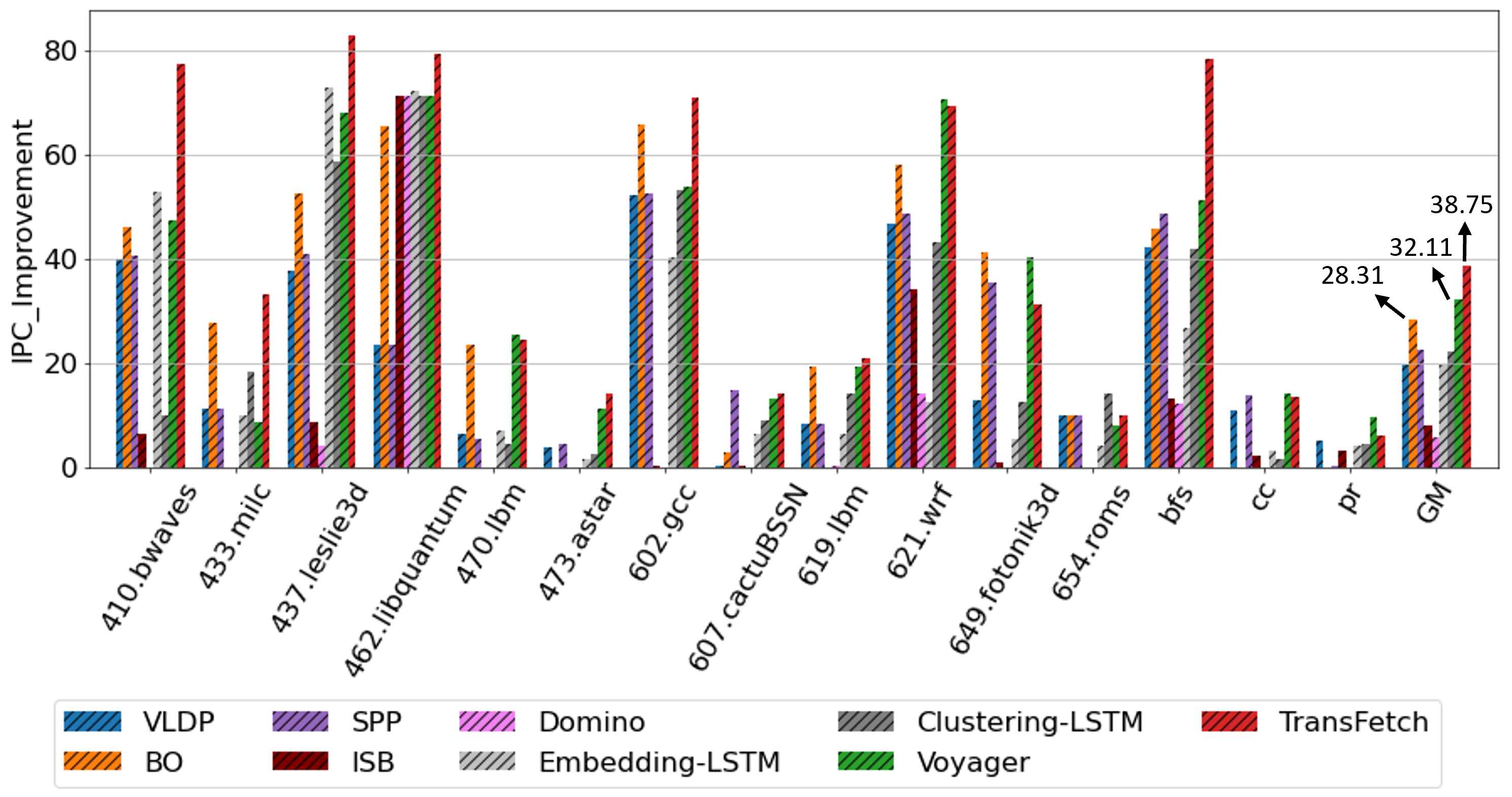}
  \caption{IPC improvement of~\ourwork~and baselines.}
  \label{fig:ipc}
\end{figure}

Figure~\ref{fig:ipc} shows the IPC improvement of the prefetchers, which indicates the overall contribution of a prefetcher to the system speedup. In the average,~\ourwork~achieves the highest IPC improvement at 38.75\%, which is 10.44\% higher than BOP and 6.64\% higher than Voyager.
VLDP, SPP, ISB, Domino, Embedding-LSTM, and Clustering-LSTM achieve lower IPC improvement at 19.50\%, 22.59\%, 8.01\%, 5.64\%, 19.46\%, and 21.99\%, respectively.


There are cases when~\ourwork~significantly outperforms other prefetchers. For \textit{410.bwaves},~\ourwork~achieves 77.44\% IPC improvement, compared with Embedding-LSTM at 52.89\% and BOP at 46.12\%. For \textit{bfs},~\ourwork~ahieves 78.5\% IPC improvement, compared with the second highest Voyager at 51.14\% and the highest rule-based prefetcher SPP at 48.8\%.



\subsection{Distance Prefetching Evaluation}

In ideal implementation, assuming full parallelism in our model, the estimated latency $T\approx 100$ cycles according to Equation~\ref{eq:latency} and Table~\ref{tab:model}, with dimensions $D=64$, layer $L$=2, 
matrix multiplication $T_{mm}=1+\log_2D$, and 1 cycle lookup table implemented activations.
Recent works have explored more efficient implementations, e.g., replacing matrix multiplication by lookup tables~\cite{razlighi2017looknn} and combinational logic~\cite{nazemi2021nullanet}. 
In future implementations, the range of $T<200$ can be a reasonable target.
\begin{figure}[h]
  \centering
  \includegraphics[width=\linewidth]{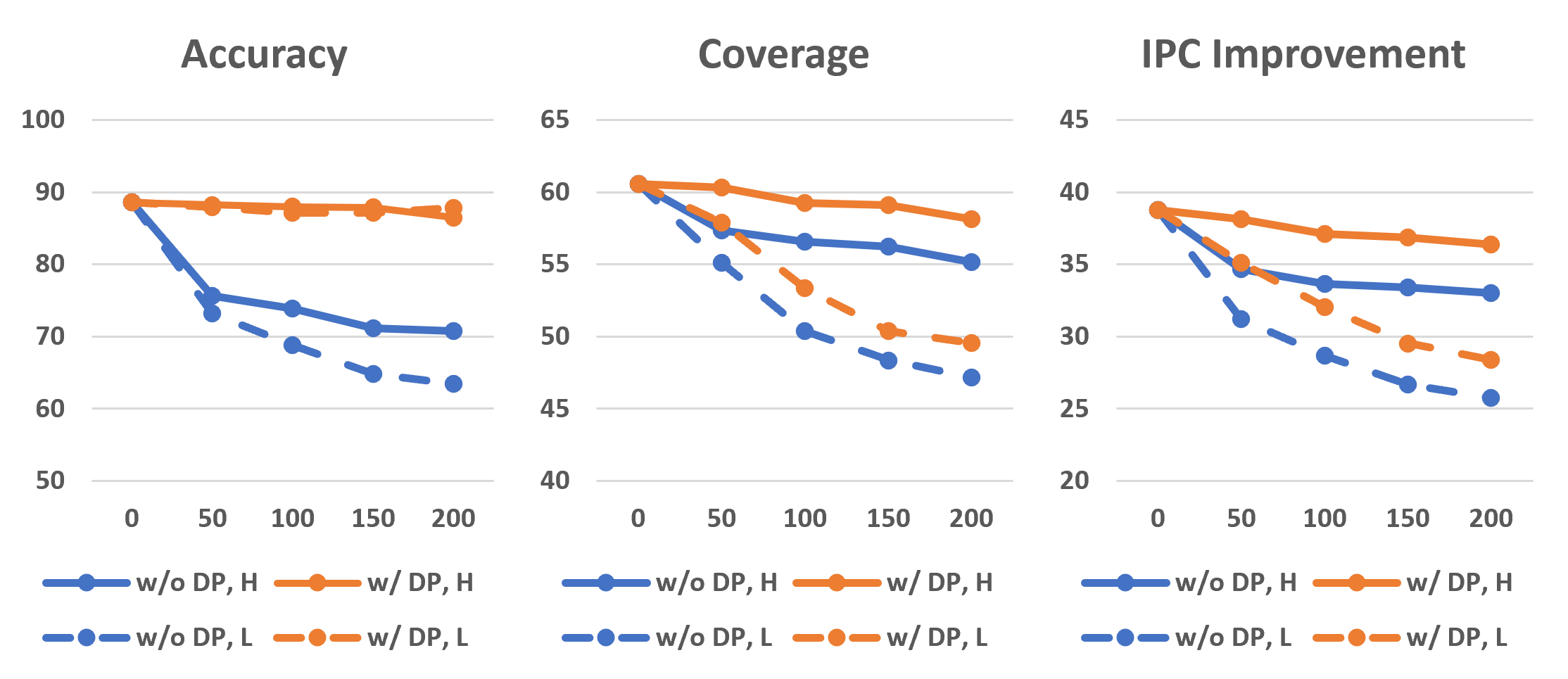}
  \caption{Effectiveness of distance prefetching. "L" means low throughput (1/T); "H" means high throughput (1).}
  \label{fig:dp}
\end{figure}

We train and simulate~\ourwork~with induced latency $T$ from 0 to 200 cycles, under bounds of throughput $1/T$ and 1 inference per cycle. The average prefetching performance is shown in Figure~\ref{fig:dp}.
With 200 cycles latency, the high throughput model with distance prefetching (DP) achieves 36.29\% IPC improvement, higher than the model without DP at 34.67\%, both are still highest compared with the baselines in Figure~\ref{fig:ipc}. 
Even for the low throughput models, DP shows the IPC improvement at 28.39\% for 200 cycles latency, slightly higher than BOP at 28.31, while the models without DP drop to 25.78\%. 
Overall, distance prefetching effectively decelerates the performance drop caused by inference latency.

\section{Conclusion}
In this paper, we presented~\ourwork, a novel way to model prefetching and to solve the problem of \textit{class explosion, tokenization, labeling}, and \textit{latency}. The keys to our approach are using fine-grained address segmentation for model input to reduce vocabulary and avoid tokenization, using delta bitmap for labeling, and using powerful and parallelizable attention-based network for prediction.~\ourwork~achieves 26\% - 36\% higher F1-score than delta inputs and 15\%- 24\% higher F1-score than page \& offset inputs.~\ourwork~achieves 38.75\% IPC improvement in simulation, outperforming the best-performing rule-based prefetcher BOP by 10.44\% and ML-based prefetcher Voyager by 6.64\%. We believe~\ourwork~offers a new paradigm for modeling prefetching toward high performance and practicality. In future work, we plan to explore the incorporation of software hints to improve prefetching performance. 

\begin{acks}
This work was supported by National Science Foundation (NSF) under award number CCF-1912680.

\end{acks}

\bibliographystyle{ACM-Reference-Format}
\bibliography{cf22}


\end{document}